\documentclass[12pt]{iopart}
\expandafter\let\csname equation*\endcsname\relax
\expandafter\let\csname endequation*\endcsname\relax
\usepackage{amsmath}
\usepackage{graphicx}
\usepackage{fancyhdr}
\usepackage{amssymb,amsfonts,color}
\usepackage[colorinlistoftodos]{todonotes}

\definecolor{red}{rgb}{1,0,0}

\begin{document}

\title[]{Ultralocal nature of geometrogenesis}

\author{Micha{\l} Mandrysz$^{a}$ and Jakub Mielczarek$^{a,b}$}

\address{$^{a}$Institute of Physics, Jagiellonian University, {\L}ojasiewicza 11, 
30-348 Cracow, Poland \\
$^{b}$Centre de Physique Th\'eorique, 163 Avenue de Luminy, F-13288 Marseille, France}

\begin{abstract}
In this article we show that the ultralocal state of gravity can be associated with the 
so-called crumpled phase of gravity, observed e.g. in Causal Dynamical Triangulations. 
By considering anisotropic scaling present in the Ho\v{r}ava-Lifshitz theory, we prove 
that in the ultralocal scaling limit ($z \rightarrow 0$) the graph representing connectivity 
structure of space is becoming complete. In consequence, transition from the ultralocal 
phase ($z=0$) to the standard relativistic scaling ($z=1$) is implemented by the
\emph{geometrogensis}, similar to the one considered in Quantum Graphity approach.  
However, the relation holds only for the finite number of nodes $N$ and in the continuous
limit ($N\rightarrow \infty$) the complete graph reduces to the set of disconnected 
points due to the weights $w=1/N$ associated with the links. By coupling Ising spin matter 
to the considered graph we show that the process of geometrogensis can be associated with 
critical behavior. Based on both analytical and numerical analysis phase diagram of the 
system is reconstructed showing that (for a ring graph) symmetry broken phase occurs 
at $z\in [0, 0.5)$. Finally, cosmological consequences of the considered process of 
geometrogenesis as well as similarities with the so-called synaptic pruning are briefly discussed.  
\end{abstract}

\maketitle

\section{Introduction}

One of the characteristic features of dynamics of gravitational field near the space-like 
singularities is the \emph{ultralocality} known as one of the pillars of the BKL conjecture 
\cite{Belinsky:1970ew,Belinsky:1982pk}. The ultralocality is associated with suppression 
of the spatial derivatives and collapse of the light cones. In consequence, spatial propagation 
of information is prohibited, which is reflected by the \emph{asymptotic silence} 
\cite{Andersson:2004wp} of the BKL dynamics. 

Besides the classical theory of gravity, the ultralocality appeared also in some attempts of
addressing quantum nature of gravitational interactions. In particular, in has been shown 
that the ultralocal phase emerges in the context of loop quantum cosmology \cite{Mielczarek:2012tn}, 
being associated with quantum deformations of space-time symmetries \cite{Bojowald:2012ux}. 
Furthermore, recent studies suggest that the ultralocal emerges also in the Causal Sets approach 
to Planck scale physics \cite{Eichhorn:2017djq}. Moreover, in this article we will employ the 
fact that the ultralocal limit can be recovered as a special case of the anisotropic scaling 
in the Horava-Lifshitz theory \cite{Horava:2009uw}. 

While the ultralocal state is characterized by the vanishing interactions between neighboring 
points, there is another phase considered in the context of gravity which seem to describe just 
an opposite situation. The phase is called \emph{crumpled} and is associated with a high
connectivity between the points of space. Such behavior has been observed in the 
so-called $B$ phase of the Causal Dynamical Triangulations (CDT) \cite{Ambjorn:2012ij}, 
which is a nonperturbative approach to quantum gravity. In the $B$ phase of CDT the valence 
of vertices of triangulation tend to infinity while valence of the graph dual to triangulation 
remains fixed. Furthermore, in the co-called Quantum Graphity \cite{Konopka:2006hu} approach 
to gravity (related to string nets approach \cite{Levin:2004mi}) the crumpled phase has 
been introduced as a high temperature pre-geometric state of gravitational field. Another
example is the emergence of cosmological configuration in the Group Field Theory 
approach which also reflects this kind of emergent behavior \cite{Oriti:2016acw}.  The 
transition from non-geometric (in particular crumpled) state to the one characterized by 
classical geometric properties in known under the name of \emph{geometrogenesis} 
(see Ref. \cite{Mielczarek:2014aka}).
  
The main objective of this article is to investigate a conjecture presented in Ref. \cite{Mielczarek:2017cdp} 
which stated that the two seemingly different phases, the ultralocal one and the crumpled one
are inherently interrelated. 

In order to explore this relation, in Sec. \ref{Horava} we consider the anisotropic scaling which 
allows us to study transition between the relativistic ($z=1$) and ultralocal ($z=0$) phases 
in a controlled way. By analyzing the corresponding Laplace operator we infer what is
the connectivity structure of underlying graph representing spatial geometry. We show that 
the graph associated with the ultralocal limit is a complete graph. However, because $1/N$ 
weights are associated with the links, the complete graph reduces to a set of disconnected 
nodes in the thermodynamic limit ($N\rightarrow \infty$). Then, in Sec. \ref{Ring} the 
special case of a ring graph is introduced. In Sec. \ref{SpectraDimension} behavior of the 
spectral dimension as a function diffusion time and $z\in [0,1]$ is investigated. 
In order to go beyond the ``gravitational'' sector, in Sec. \ref{Ising} a toy model of the matter 
content in the form of Ising spins is introduced. The spins are defined at the nodes of the 
graph. The phase diagram of the system for the ring graph is reconstructed 
using both analytical and numerical methods. We show that the system displays critical 
behavior, and the symmetry broken phase occurs for $z\in [0, 0.5)$. In Sec. \ref{Cosmo} 
general discussion concerning cosmological significance of the discussed process of 
geometrogenesis is given.  In particular, the issues of generation of primordial inhomogeneities 
and topological defects are stressed. In Sec. \ref{Summary} the obtained results are summarized 
and complemented with final discussions. Somewhat beyond the main subject of this article, we point 
out the striking similarity of the gravitational process described in this article with the 
so-called \emph{synaptic pruning} known to play an important role in the maturing brain.

\section{Ultralocal=Crumpled for finite $N$} \label{Horava}

The aim of this section is to prove that the ultralocal state can be associated with 
the crumpled state of space. The ultralocality is defined here such that the algebra 
of gravitational constraints takes the ultralocal form, \emph{i.e.} \cite{Isham:1975ur}
\begin{eqnarray}
\left\{D[M^a_1], D[M^a_2]\right\} &=& D[M_1^b \partial_b M^a_2 - M_2^b \partial_b M^a_1]\,, \nonumber\\
\left\{S[M], D[M^a]\right\} &=& -S[ M^a \partial_a M]\,, \nonumber\\
\left\{S[M_1], S[M_2]\right\} &=& 0, 
\label{HDA}
\end{eqnarray}
where $S[M]$ and $D[M^a]$ are scalar and diffeomorphism constraints correspondingly.  
The constraints are parametrized by the lapse function $M$ and the shift vector $M^a$. 
In contrast to the General Relativistic (GR) hypersurface deformation algebra, the algebra 
(\ref{HDA}) is a Lie algebra. There are various ways to pass to the regime where symmetry 
of the gravitational field is approximated by the algebra (\ref{HDA}). At the dynamical level, 
such limit is recovered e.g. in the mentioned BKL scenario or as a special case of the loop 
quantum deformations of the algebra of constraints \cite{Bojowald:2012ux}. The algebra 
(\ref{HDA}) is also recovered in the strong coupling limit ($G_{N} \rightarrow \infty$)  
of the gravitational field. Yet another possibility, which we employ here is related with the 
anisotropic space-time scaling introduced in the Horava-Lifshitz theory \cite{Horava:2009uw}.  

While relativistically both space and time components scale in the same way, in the Horava-Lifshitz 
approach departure from such a case is allowed. The  corresponding anisotropy of space-time scaling 
is parametrized by the exponent $z$ introduced as follows \cite{Horava:2009uw}:  
\begin{equation}
{\bf x} \rightarrow b {\bf x}, \ \ \ t \rightarrow b^z t, \label{scalling}
\end{equation}
where $b$ is some positive definite constants. In the $z=1$ case the standard relativistic space-time 
scaling is recovered. On the other hand, for $z=3$ it has been shown that the corresponding theory 
of gravity is characterized by power-counting renormalizability \cite{Horava:2009uw}. It has also 
been shown that in the $z\rightarrow 0$ limit the theory of gravity satisfies the ultralocal algebra 
(\ref{HDA}) \cite{Horava:2009uw}. 

One of the important consequences of the anisotropic scaling (\ref{scalling}) is that form of the 
Laplace operator is $z-$dependent. Namely, in the case of general $z$, the standard Laplace 
operator $\Delta :=- \nabla^2$ (keep in mind the minus sign convention) generalizes to \cite{Horava:2009if}:
\begin{equation}
\Delta_z := \Delta^z, \label{Deltaz}
\end{equation} 
which is defined such that ellipticity of the operator  is guaranteed (the eigenvalues will 
remain positive definite). Worth mentioning here is that the $\Delta_z$ operator for $z\in(0,1)$
can be interpreted as a fractional Laplacian \cite{Kwasnicki}. The $\Delta$ is a standard Laplace 
operator defined on either Euclidean $d$-dimensional space ($\mathbb{R}^d$) or a graph. Here, we will focus on 
the case of the operator $\Delta$ defined on a graph characterized by adjacency 
matrix $A$ and degree matrix $D$. Physically, the graph is representing relational 
structure between the points of space\footnote{In the Planck scale picture of space we 
can think about the points as elementary chunks (atoms) of space.}. We will denote 
the number of nodes (atoms of space) of the graph as $N$. With the use of the $A$ and 
$D$ matrices, the Laplace operator $\Delta$ can be expressed as $\Delta = D-A$.

By considering the new Laplace operator (\ref{Deltaz}) one can now ask about the effective
graph associated with this operator. In other words, one can consider the usual decomposition:
\begin{equation}
\Delta_z  =D_z-A_z,
\end{equation}
and ask what are the forms of the matrices $D_z$ and $A_z$, knowing the initial $A$ and 
$D$ used to define $\Delta_z$.  In what follows, we will reconstruct the form of both 
$A_z$ and $D_z$ in the ultralocal ($z\rightarrow 0$) limit.   
 
For this purpose, let us now assume that $z=\frac{1}{n}$, with $n\in \mathbb{N}$ and 
study the $n\rightarrow \infty$ ($z\rightarrow 0$) limit. Consistency requires that 
\begin{equation}
\Delta=\Delta^{n}_z \label{Delta}.
\end{equation} 

The further steps require eigendecomposition of the Laplacian matrix:
\begin{equation}
\Delta = P \Lambda P^{-1} \label{decompDelta}
\end{equation} 
where $\Lambda={\rm diag}(\lambda_1, \dots, \lambda_N)$ is a diagonal 
$N\times N$ matrix. The $\lambda_n$ are eigenvalues of the Laplacian matrix $\Delta$,
corresponding to the eigenvalue equation $\Delta e_n = \lambda_n e_n$ and $e_n$
are the eigenvectors. The $P=(e_1, e_2,\dots, e_n)$ is a transformation matrix, which 
in the considered case of symmetric and real matrix $\Delta$ satisfies the hermiticity
condition $P P^{\dagger} = P^{\dagger} P  = \mathbb{I}$. Furthermore, one can 
show that the eigenvectors $e_n$ are orthonormal, i.e. $e_n^{\dagger} e_m =\delta_{nm}$. 
An important property is that for the Laplace operator the eigenvalues can be arranged 
in a non-decreasing order $0  = \lambda_1 \leq \lambda_2 \leq \dots \leq \lambda_N$. 
The number of zero eigenvalues is known to be directly related to the number 
of disconnected components of the graph. Here, without loss of generality, the 
simplest case of a connected graph is considered, such that there is only one zero 
eigenvalue and $\lambda_2 > \lambda_1=0$. Because disconnected graphs are 
represented by sub-square matrices in a larger matrix, the study of properties of 
disconnected graphs can be reduced to the study of connected graphs. In the considered 
case the zero eigenvalue corresponds simply to the trivial eigenvector
$e_1 = (1/\sqrt{N},1/\sqrt{N},\dots,1/\sqrt{N})^T$, 
normalized such that $e_1^{\dagger} e_1 =1$.  In this case, the eigenequation 
reduces to  $\Delta e_1 = \lambda_1 e_1=0$. 

One can now observe that the relation $\Delta^{\frac{1}{n}} = P \Lambda^{\frac{1}{n}} P^{-1}$ 
must be satisfied since the $n$-th power of this equation correctly reduces to Eq. \ref{decompDelta}.
Based on this, together with definition (\ref{Deltaz}), one can express $\Delta_z$ as:
\begin{equation}
\Delta_z =  P \Lambda^z P^{-1}. \label{DeltazDecomp}
\end{equation} 
This equation allows us to study the (ultralocal) $z\rightarrow 0$ limit:
\begin{eqnarray}
\left[\Delta_0\right]_{ij} &: =& \lim_{z\rightarrow 0}\left[\Delta_z\right]_{ij} = \left[P \left( \lim_{z\rightarrow 0}   
\Lambda^z \right) P^{-1}\right]_{ij} = \left[P {\rm diag}(0,1,  \dots, 1) P^{\dagger}\right]_{ij} \nonumber \\
              &=& \sum_{k=2}^{N} P_{ik}P^{\dagger}_{kj} = \delta_{ij} -P_{i1}P^{\dagger}_{1j}.  \label{Delta0deriv}
\end{eqnarray}
With the use the fact that the $P_{i1}$ column of the transformation matrix $P$ is built from the
$e_1 = (1/\sqrt{N},1/\sqrt{N},\dots,1/\sqrt{N})^T$ eigenvector and correspondingly $P_{1j}^{\dagger}$ is built 
from $e_1^{\dagger} = (1/\sqrt{N},1/\sqrt{N},\dots,1/\sqrt{N})$, one can reduce (\ref{Delta0deriv}) to
\begin{equation}
\left[\Delta_0\right]_{ij} =\delta_{ij}-\frac{1}{N}=\left[D_0\right]_{ij}-\left[A_0\right]_{ij},    \label{UltralocalDelta}
\end{equation}
where $A_0$ and $D_0$ are respectively adjacency and degree matrices 
corresponding to the Laplace operator $\Delta_0$.  The components of $A_0$ 
and $D_0$ take the following forms:
\begin{eqnarray}
\left[A_0\right]_{ij}  &=& (1-\delta_{ij}) w,  \label{A0} \\
\left[D_0\right]_{ij} &=& \delta_{ij}(N-1) w, \label{D0}  
\end{eqnarray}
where the weights $w_{ij}=w=\frac{1}{N}$ for $i\neq j$. The associated graph is of the 
complete type and may describe crumpled state of gravity. The only difference with respect 
to the usually considered complete graph is that the links are now weighted with $w = \frac{1}{N}$. 
Since all points are connected, the notion of spatial extension breaks down - space is effectively 
becoming a single point. This view provides one with intuitive explaination of why 
ultralocality is associated with the crumpled phase. 
On the other hand, in the continuous (thermodynamic) limit ($N\rightarrow \infty$) 
the complete graph reduces to the set of disconnected points due to the weights $w=1/N$ 
associated with the links. In this limit the graph takes the form which is usually associated 
with the ultralocal phase and is characterized by the following adjacency and degree matrices: 
\begin{eqnarray}
\lim_{N\rightarrow \infty} \left[A_0\right]_{ij}  &=& 0,  \label{A0cont} \\
\lim_{N\rightarrow \infty} \left[D_0\right]_{ij} &=& \delta_{ij}. \label{D0cont}  
\end{eqnarray}

\begin{figure}[ht!]
\centering
\includegraphics[width=10cm]{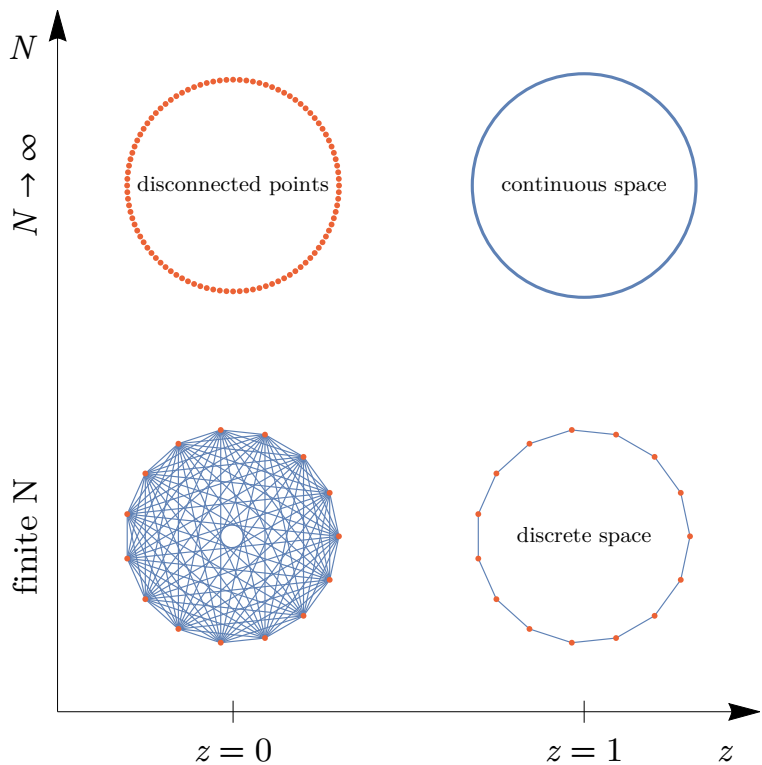}
\caption{Pictorial representation of the phases of connectivity as a function of 
the anisotropic scaling $z$ and the number of nodes $N$. For finite $N$, 
the exemplary ring graph (bottom right) at $z=1$ transforms into complete 
graph (bottom left) in the ultralocal limit ($z\rightarrow 0$). In the continuous 
 ($N\rightarrow \infty$) limit the ring graph reduces to a circle (top right) and 
 the complete graph transfigures into a set of disconnected points (top left).}
\label{GG}
\end{figure}

In Fig. \ref{GG} we present the dependence of the structure of connectivity on example of a ring graph.
For finite $N$, the ring graph at $z=1$ transitions into a complete graph 
in the ultralocal limit $z\rightarrow 0$). While performing the continuous 
($N\rightarrow \infty$) limit the graph at $z=1$ reduces to a circle ($S^1$) 
and a complete graph at $z=0$ reduces to a set of disconnected points. 
The later behavior is due to the weights $w = \frac{1}{N}$ associated with 
the links in the ultralocal limit. Worth mentioning here is that the three 
configurations: discrete space, disconnected points and the crumpled state
have also been observed to occur together in the Quantum Graphity approach 
\cite{Wilkinson:2014zga}.

The connected graph (at $z=0$ and finite $N$) represents a crumpled phase, 
in particular similar to the one observed in CDT. To make the relation with CDT 
closer one can view the considered graphs as graph of nodes of triangulations 
in CDT. The observations made above suggest that the $B$ phase of CDT 
(observed for both spherical \cite{Coumbe:2015oaa} and toroidal \cite{Ambjorn:2018qbf} 
topologies) reflects the ultralocal state of gravity. Because in the CDT simulations, a finite 
number of simplices ($N$) is considered the phase manifests as the crumpled state.   

One of the consequences of the observation made here is that puzzling divergence 
of the spectral dimension in the ultralocal limit predicted by  Horava-Lifshitz theory 
can now be understood. This issue is addressed in Sec. \ref{SpectraDimension}. 

\section{The ring graph model} \label{Ring}

Results of the previous section are general and are not dependent on dimensionality 
of the universe (graph) in the $z=1$ limit. However, it would be difficult to analyze further
properties of the geometrogenesis without making assumptions about the relational
structure in the geometric limit. The case which allows for a broad analytical 
analysis is the $S^1$ geometry, which in discrete version is described by a ring graph. 
Therefore, in our further investigations we will focus on a model in which the $z=1$ 
limit is described by a ring graph composed of $N$ nodes. One can perceive this
case as a prototype low-dimensional model of a spatially compact universes. 

The ring graph is invariant under cyclic permutation of indices $i \rightarrow i+1$, 
associated with the $\mathbb{Z}_N$ group. The symmetry is generated by the 
group element $g$, which commutes with the Laplace operator $\Delta$. Because 
of this, $\Delta$ and $g$ have the common set of eigenvectors. Using the fact that 
$g^N=\mathbb{I}$, together with the eigenequation $g e_n = g_n e_n$ one can 
show that the eigenvalues $g_n$ correspond to the complex roots of the equation 
$g_n^N=1$, which are $g_n=e^{i 2\pi (n-1)/N}$ and $n=1,2,\dots, N$. Now, with the 
use of the action of the group element $g$ on the eigenvectors $g (e_n)_j = (e_{n})_{j+1}$, 
functional form of the normalized eigenvectors can be found:
\begin{equation}
(e_n)_j = \frac{1}{\sqrt{N}} e^{i 2\pi (n-1)j/N},     \label{eigenfunctions}
\end{equation}
which are normalized using $\sum_{j=1}^N (e_n)_j (e_m^*)_j  =\delta_{nm}$. 
Applying the derived eigenfunctions to the eigenequation $\Delta e_n = \lambda_n e_n$, 
the eigenvalues of the Laplace operator  $\Delta$ can be determined: 
\begin{equation}
\lambda_n = 2\left[1-\cos\left( 2\pi \frac{(n-1)}{N} \right) \right],  \label{EVCircle Graph}
\end{equation}
where $n=1,2,\dots, N$. The eigenvalues applied to Eq. \ref{DeltazDecomp} allow 
us to determine eigenvalues of the $\Delta_z$ Laplace operator, which are elements of the 
diagonal matrix:
\begin{equation}
\Lambda^z =  {\rm diag}(\lambda_1^z,\lambda_2^z,  \dots, \lambda_N^z), 
\end{equation}
where $\lambda_n$ are given by Eq.  \ref{EVCircle Graph}. In consequence, one 
write eigenvalues of the $\Delta_z$ Laplace operator as:
\begin{equation}
\lambda^z_n = 2^z\left[1-\cos\left( 2\pi \frac{(n-1)}{N} \right) \right]^z,  \label{EVzCircle Graph}
\end{equation}
where $n=1,2,\dots, N$. Furthermore,  because $\Delta_z =  P \Lambda^z P^{\dagger}$ and the 
$P$ matrices can be expressed in terms of the eigenfunctions (\ref{eigenfunctions}) 
the matrix elements of $\Delta_z$ can be written as follows:
\begin{align}
\left( \Delta_z \right)_{mn} &= P_{m k} \Lambda_{k l}^z P_{l n}^{\dagger} = 
\sum_{k=1}^{N} \sum_{l=1}^{N} (e_k)_m \delta_{kl} \lambda^z_{k}(e^*_l)_n =  
\frac{1}{N}\sum_{k=1}^{N}  e^{i 2\pi (k-1)(m-n)/N}  \lambda^z_{k}      \nonumber \\ 
&=  \frac{1}{N}\sum_{k=1}^{N}  e^{i 2\pi (k-1)(m-n)/N} 2^z
\left[1-\cos\left( 2\pi \frac{(k-1)}{N} \right) \right]^z. \label{DeltazComplex}
\end{align}
Furthermore, the $\Delta_z$ is a real matrix (proof can be found in \ref{realMatrixAppendix}). Therefore, the imaginary component of Eq.  (\ref{DeltazComplex})
must vanish and we can write that 
\begin{align}
\left( \Delta_z \right)_{mn} =  \frac{1}{N}\sum_{k=1}^{N}  \cos \left[ 2\pi (k-1)(m-n)/N\right] 2^z\left[1-\cos\left( 2\pi \frac{(k-1)}{N} \right) \right]^z. 
\label{DeltazReal}
\end{align}
It is straightforward to show that the  ultralocal limit (\ref{UltralocalDelta}) is correctly recovered from 
this expression. The valence of a $n$-th node is now easy to determine from:
\begin{equation}
d_{n} := \left( \Delta_z \right)_{nn} = \frac{2^z}{N} \sum_{k=1}^{N}\left[1-\cos\left( 2\pi \frac{(k-1)}{N} \right) \right]^z.    \label{dn}
\end{equation}
In particular, in the ultralocal limit ($z\rightarrow 0$) we obtain 
\begin{equation}
\lim_{z\rightarrow 0} d_{n}=\frac{1}{N} \sum_{k=2}^{N}1 = \frac{1}{N} (N-1),  \label{dnzzero} 
\end{equation}
in accordance with Eq. \ref{D0}. The $1/N$ factor corresponds to the weight $w$ associated with the 
links and the $N-1$ factor counts the number of links attached to any node. 

\begin{figure}[ht!]
\centering
\includegraphics[width=10cm, angle = 0]{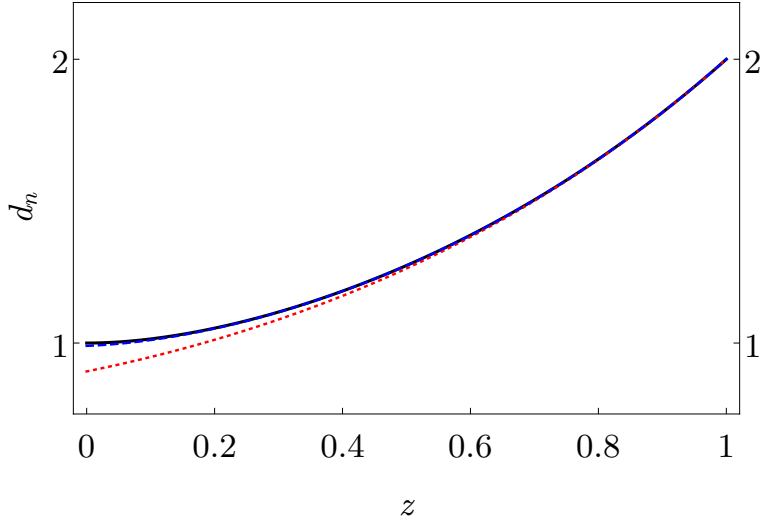}

\caption{Effective degree of nodes $d_n$ as a function of $z$. The (black) continuous curve corresponds 
to the $N\rightarrow \infty$ case given by Eq. \ref{dnthermo}. The (red) dotted curve plotted for $N=10$ 
and the (blue) dashed curve corresponds to $N=100$ and overlaps with the curve for $N\rightarrow \infty$.}
\label{valence}
\end{figure}

It is interesting to observe that, the thermodynamic limit  ($N\rightarrow \infty$) of the expression (\ref{dn}) can 
be performed by introducing the angle variable $\theta := 2\pi \frac{(k-1)}{N}$. This allows us to write: 
\begin{equation}
d_{n}^{N\rightarrow \infty} :=\lim_{N\rightarrow \infty} d_{n} = \frac{2^z}{2 \pi } \int_0^{2 \pi } (1-\cos{\theta} )^z d \theta 
= \frac{4^z \Gamma  \left(z+\frac{1}{2}\right)}{\sqrt{\pi } \Gamma  (z+1) } < \infty, \label{dnthermo}
\end{equation}
for $z \in [0,1]$, where $\Gamma(x)$ is the gamma function. In Fig. \ref{valence} we plot behavior of the effective 
degree of nodes  given by Eq. \ref{dn} and Eq. \ref{dnthermo}.
The value of $2$ is correctly recovered in for $z=1$. On the other hand, in the ultralocal limit ($z\rightarrow 0$)
the value of $d_n$ converges to $d_n=1$ as expected based on Eq. \ref{dnzzero} taken in the $N\rightarrow \infty$ limit. 

\section{Spectral dimension} 
\label{SpectraDimension}

In the recent years the so-called \emph{spectral dimension} attracted significant 
attention in analyzes of different quantum gravitational space(time) configurations. 
The studies concerned such approaches to quantum gravity and models of the 
Planck scale physics as Causal Dynamical Triangulations \cite{Ambjorn:2005db}, 
Loop Quantum Gravity \cite{Mielczarek:2016zfz}, Multifractal theories \cite{Calcagni:2016azd}, 
Asymptotic Safety \cite{Lauscher:2005qz}, Causal Sets \cite{Belenchia:2015aia}, 
Deformed Poincar\'e algebras \cite{Arzano:2014jfa}, etc.  

The spectral dimension is defined with the use of diffusion process performed in 
some external time $\sigma$. The idea behind the definition of the spectral dimension 
is based on the diffusion time dependence of the average return probability $P(\sigma)$ 
which can be determined as a trace of the Heat Kernel $K({\bf x}, \tau, {\bf x}', \tau'; \sigma )$. 

In the considered case of the anisotropic scaling (\ref{scalling}), employing the 
appropriate spatial Laplace operator (\ref{Deltaz}), the heat kernel equation takes 
the following form \cite{Horava:2009if}:
\begin{equation}
\frac{\partial}{\partial \sigma} K({\bf x}, \tau, {\bf x}', \tau'; \sigma ) = \left(\frac{\partial^2}{\partial \tau^2}
-\Delta_z  \right)K({\bf x}, \tau, {\bf x}', \tau'; \sigma ), \label{HKEz}
\end{equation}
here $\tau = -i t $ is the Wick rotated coordinate time $t$. Continuous solution to
 (\ref{HKEz}) can be found and the corresponding value of the spectral 
dimension can be determined \cite{Horava:2009if}:
\begin{equation}
d_S := -2 \frac{d \log P(\sigma)}{d \log \sigma} = 1+\frac{d}{z}, \label{SDHL}
\end{equation}
where $d$ is the spatial topological dimension of the spatial part. 

In case of GR ($z=1$), the spectral dimension (\ref{SDHL}) reduces to the space-time 
dimension $d_S=1+d$. However, in the relevant ultralocal limit $z\rightarrow 0$, the 
spectral dimension tends to infinity, $d_S \rightarrow \infty$. As we will show below, 
such behavior is expected for the complete graph describing the ultralocal phase
for a finite $N$. 

For our purposes it is relevant to consider diffusion without the time dimension, which 
can be easily added to the final results. 
Then, for the matrix Laplace operator $\Delta_z$ the solution to the 
Heat Kernel equation can be expressed in terms of eigenvectors $e_n$ and eigenvalues 
$\lambda^z_n$ of the operator. Straightforward analysis leads to the following expression:
\begin{equation}
K(i,j;\sigma) = \sum_{n=1}^N e^{-\lambda^z_n \sigma} (e_{n})_i (e^{*}_{n})_j,
 \end{equation}
which allows to derive expression of the average return probability 
\begin{equation}
P(\sigma):= {\rm tr } K(i,j;\sigma) = \sum_{n=1}^N e^{-\lambda^z_n \sigma}.
\end{equation}
Based on the above expression the spectral dimension can be written as follows:
\begin{equation}
d_{S}= 2\sigma \frac{\sum_{n=1}^N \lambda^z_n e^{-\lambda^z_n \sigma}}{\sum_{n=1}^N e^{-\lambda^z_n \sigma}}.  
\label{SDlambdaz}
\end{equation}

\begin{figure}[ht!]
\centering
\includegraphics[width=10cm]{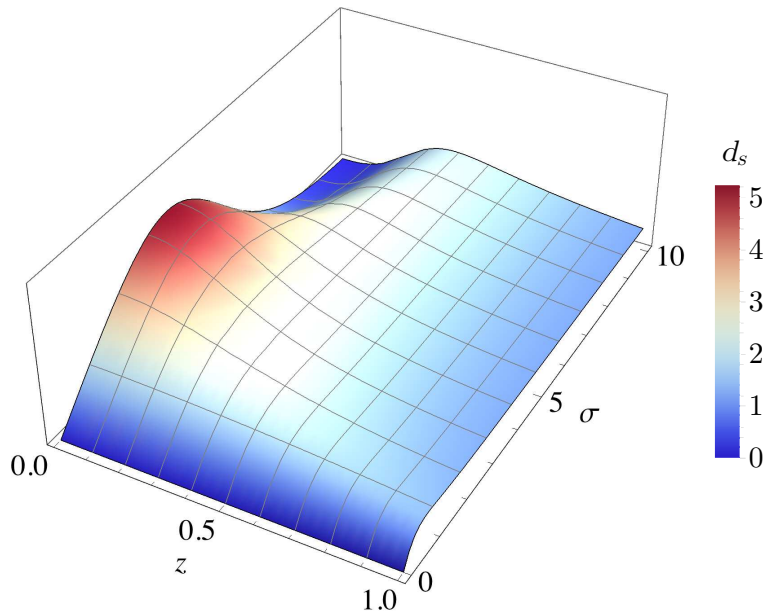}
\caption{Spectral dimension as a function of diffusion time $\sigma$ and the 
anisotropic scaling parameter $z$ for a ring graph with $N=100$. For $z=1$ the 
spectral dimension saturates $d_S=1$ as expected for the ring graph. In the 
ultralocal limit, the spectral dimension is peaked at some $\sigma_* \sim 1/N$ 
and falls to zero for $\sigma \rightarrow \infty$.}
\label{SD}
\end{figure}

In Fig. \ref{SD} we plot Eq.  \ref{SDlambdaz} as a function of both diffusion time $\sigma$ and the 
anisotropic scaling parameter $z$ for a ring graph with $N=100$.

In the case of a complete graph which corresponds to the $z\rightarrow0$ phase 
the besides the zero eigenvalue $\lambda_{1}=0$ all remaining eigenvalues 
are equal and given by $\lambda_{2}=\lambda_{3}=\dots = \lambda_{N}=1$. 
In consequence, the analytic expression for the spectral dimension is \cite{Mielczarek:2014aka}
\begin{equation}
d_{S} = \frac{2 \sigma (N-1)}{e^{\sigma}+(N-1)}.
\end{equation}
The spectral dimension peaks at some $\sigma_* \sim 1/N$ and falls to zero for 
$\sigma \rightarrow  \infty$. In the continuous ($N\rightarrow \infty$) limit 
$d_{S}\rightarrow \infty$ at $\sigma =0$ and is equal zero elsewhere. The 
divergence of the spectral dimension in the ultralocal limit is therefore a residue of 
the high connectivity of the crumpled phase. Namely, at small diffusion times there 
is a huge number of possible nodes to go, such that the corresponding dimensionality 
is perceived to be very high. 

\section{Ising matter and phase transitions} \label{Ising}

So far we have been focused solely on properties of the graph representing 
connectivity structure of space(time). The easiest way to go beyond this level 
and to introduce some local degrees of freedom is by assigning classical spins 
$s_i\in \{-1,1 \}$ to every node.  Such degrees of freedom can either be consider 
as a prototype of gravitational degrees of freedom or as a matter sector. The 
simplest nontrivial Hamiltonian for the spins is 
\begin{equation}
E=-\frac{1}{2}\sum_{i, j}w_{ij} s_i s_j- h\sum_i  s_i,  
\end{equation}
where $w_{ij}$ are the weights associated with given links and $h$ is some external bias. 

The case of $z=0$ for which $w_{ij}=\frac{1}{N}(1-\delta_{ij})$, is especially worth 
considering here and leads to
\begin{equation}
E=-\frac{1}{2N} \sum_{i\neq j}s_i s_j- h\sum_i  s_i, 
\end{equation}
the well known Curie-Weiss model. The weights introduce the $1/N$ factor in a natural way 
which correctly ensures existence of the thermodynamic limit.  The  corresponding statistical model 
can be solved analytically with the use of following observation
\begin{equation}
\left( \sum_{i} s_i \right)^2 = N+\sum_{i\neq j}s_i s_j,
\end{equation}
and by applying the Hubbard-Stratanovich transformation. Finally, one gets an equation of state 
for magnetization which predicts existence of a second order phase transition. For $z=1$ 
the phase properties depend on the effective dimensionality of the graph. In particular, in 
the one dimensional case (ring graph) no phase transition is expected while in the higher dimensional 
cases second order phase transitions are known to occur. Hence, we pose an interesting question 
- do phase transitions occur on the whole $z=[0,1)$ interval? In what follows, for simplicity, we consider
the case with $h=0$.

\subsection{Analytical results}

\begin{figure}[ht!]
\centering
\includegraphics[width=10cm]{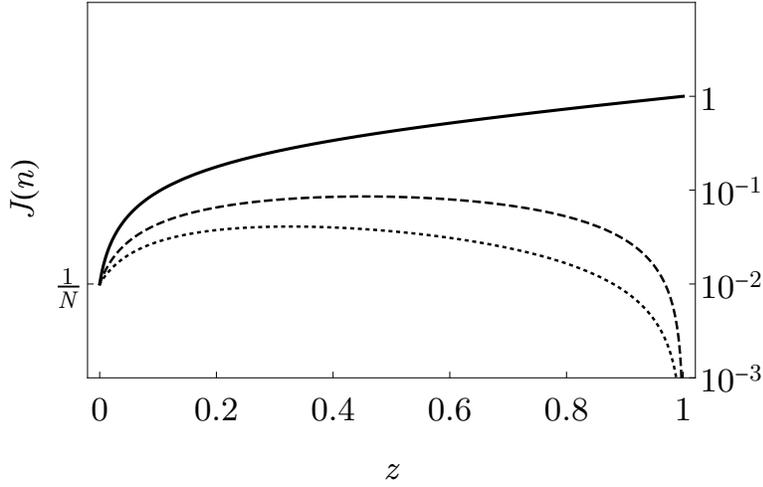}
\caption{Exemplary couplings $J(n)$ as a function of the anisotropic scaling parameter $z$ for 
a ring graph with $N=100$. The solid curve corresponds to $J(1)$, dashed to $J(2)$ and dotted to $J(3)$.}
\label{JnPlot}
\end{figure}

To answer this question we first go back to our $z$-dependent Laplacian $\Delta_z$ based on the ring graph
and determine the coupling $J(n)$ between the spin at distance $n$.  With the use of Eq. \ref{DeltazReal}
we define 
\begin{equation}
J(n) := w_{1,n+1} = -(\Delta^z)_{1,n+1}=  -\frac{2^z}{N}\sum_{k=1}^{N}  
\cos \left[ 2\pi\frac{(k-1)n}{N}\right] \left[1-\cos\left( 2\pi \frac{(k-1)}{N} \right) \right]^z.
\label{Jndisc}
\end{equation}
The important property, of this function resulting from the symmetry of the ring graph is that 
$J(i)=J(N-i)$ for all $i$. In Fig. \ref{JnPlot} we plot Eq. \ref{Jndisc}  as a function of $z$ for a ring 
graph with $N=100$.

\begin{figure}[ht!]
\centering
\includegraphics[width=10cm, angle = 0]{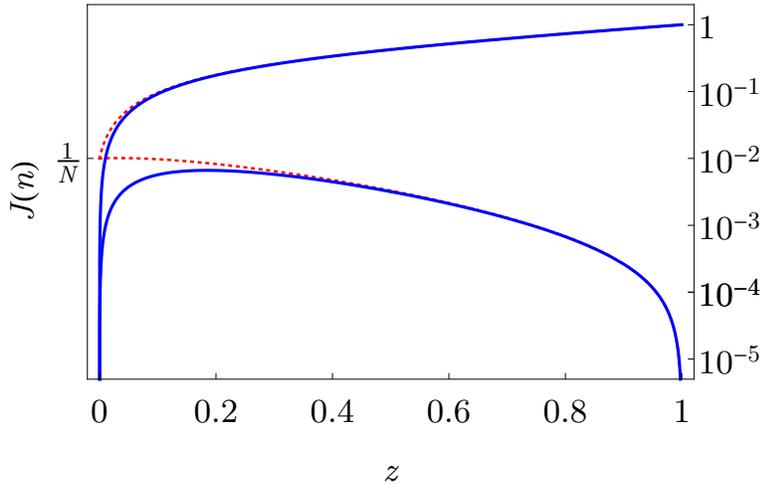}
\caption{Exemplary thermodynamical limits of $J(n)$ for $n=1$ (top curve) and $n=10$ (bottom curve) plotted 
as a function of  $z$. The dotted curves represent corresponding expression for $J(n)$ for finite $N=100$.}
\label{Jncont}
\end{figure}

For $z=1$, only $J(1)$ contributes as expected for the nearest neighbor Ising model. On the other hand, in the 
ultralocal limit ($z\rightarrow 0$) all couplings have the same value $J(n)=w=\frac{1}{N}$, in agreement with Eq. \ref{A0}.
In the thermodynamical limit ($N\rightarrow \infty$) the expression for $J(n)$ can be written as 
\begin{equation}
J(n) = -\frac{2^z}{2 \pi} \int_0^{2\pi}(1-\cos{\theta})^z \cos(\theta n)d \theta  = 
\frac{ (-1)^n \Gamma (n-z) \sin (\pi (n-z))}{\sin (2\pi z)\Gamma  (-2 z) \Gamma (1+n+z)}. \label{Jnthermo}
\end{equation}
One can notice that the interaction $J(n)$ disappears as $n \rightarrow \infty $ for $\forall z$. Thus, graph vertices distributed far away do not interact. In Fig. \ref{Jncont} we graphically compare $J(1)$ and $J(10)$ obtained with the 
use of formula (\ref{Jnthermo}) with the corresponding finite $N$ results for $N=100$. As expected, the values of $J(n)$ at $z=0$ tend to zero in the thermodynamic limit for $\forall n$. 

The purpose of investigating $J(n)$ was that the existence of phase transitions is guaranteed by fulfillment 
of the two following conditions \cite{Kubo}:
\begin{enumerate}
\item The total interaction between vertex and all of its neighbors is bounded, i.e. 
\begin{equation}
M_0 = \sum_{n=1}^{\infty} J(n) < \infty.  \label{M0}
\end{equation} 
\item The sum of interactions multiplied by their distance $n$ should be unbounded, i.e.
\begin{equation}
M_1 = \sum_{n=1}^{\infty} n J(n) = \infty. \label{M1}
\end{equation}
\end{enumerate}

The condition (i) is satisfied automatically given that double of graph interaction in one direction is always 
given by the previously found vertex degree. We verify this fact with simple calculation performed in the 
thermodynamic limit:
\begin{equation}
\begin{split}
2 M_0 & = 2 \sum_{n=1}^{\infty } J(n) = 2 \sum _{n=1}^{\infty} \frac{ (-1)^n \Gamma \left(n-z\right) 
\sin{\pi  (n-z)}}{ \sin (2 \pi z) \Gamma(-2 z) \Gamma (1+n+z) } \\ &=- \frac{\Gamma (1-z) }{2 z \cos (\pi z) \Gamma (-2z) \Gamma (1+z)} = 
\frac{4^z \Gamma  \left(z+\frac{1}{2}\right)}{\sqrt{\pi } \Gamma  (z+1) }.
\end{split}
\end{equation}

The second (ii) condition, first derived by Ruelle \cite{Ruelle} for spins arranged on infinite line, must be modified to account for the ring topology. 
The physical motivation behind this condition and its modification is simple yet intriguing, however in order to keep our reasoning short and clear we defer it to \ref{ModifiedRuelleAppendix}. The modification of Ruelle's condition results in:
\begin{align}
M_1 = a_1+\sum_{m=1}^{N/2} b_m,
\end{align}   
where 
\begin{equation}
b_m = \frac{2^{2z}}{N} \left[ \sin \left( \pi \frac{2m-1}{N} \right)\right]^{2(z-1)}. 
\end{equation}
The convergence properties (when $N\rightarrow \infty$) of the above sum can be evaluated in the continuous thermodynamical limit which can be written as
\begin{equation}
\label{M1solution}
M_{1}= \delta(z)+\frac{2^z}{2\pi} \int_0^{\pi} d\theta (\sin \theta)^{2(z-1)} =  \left\{  
\begin{array}{cc}
\infty  & \text{for}\ z \in [0,1/2]  \\
\frac{2^z}{2\sqrt{\pi}} \frac{\Gamma(-1/2+z)}{\Gamma(z)} & \text{for}\ z \in (1/2,1]  
\end{array} \right. .
\end{equation}
This equation indicates that a phase transition can occur at  $z \in [0,1/2]$. One might ask what happens with the critical temperatures in this critical segment. In the next section we show that, with limited success (partly because of finite-size effects described in \ref{FiniteSizeAppendix}) this question can be answered through numerical simulations.

\subsection{Numerical results}

\begin{figure}[ht!]
\centering
\includegraphics[width=7.7cm]{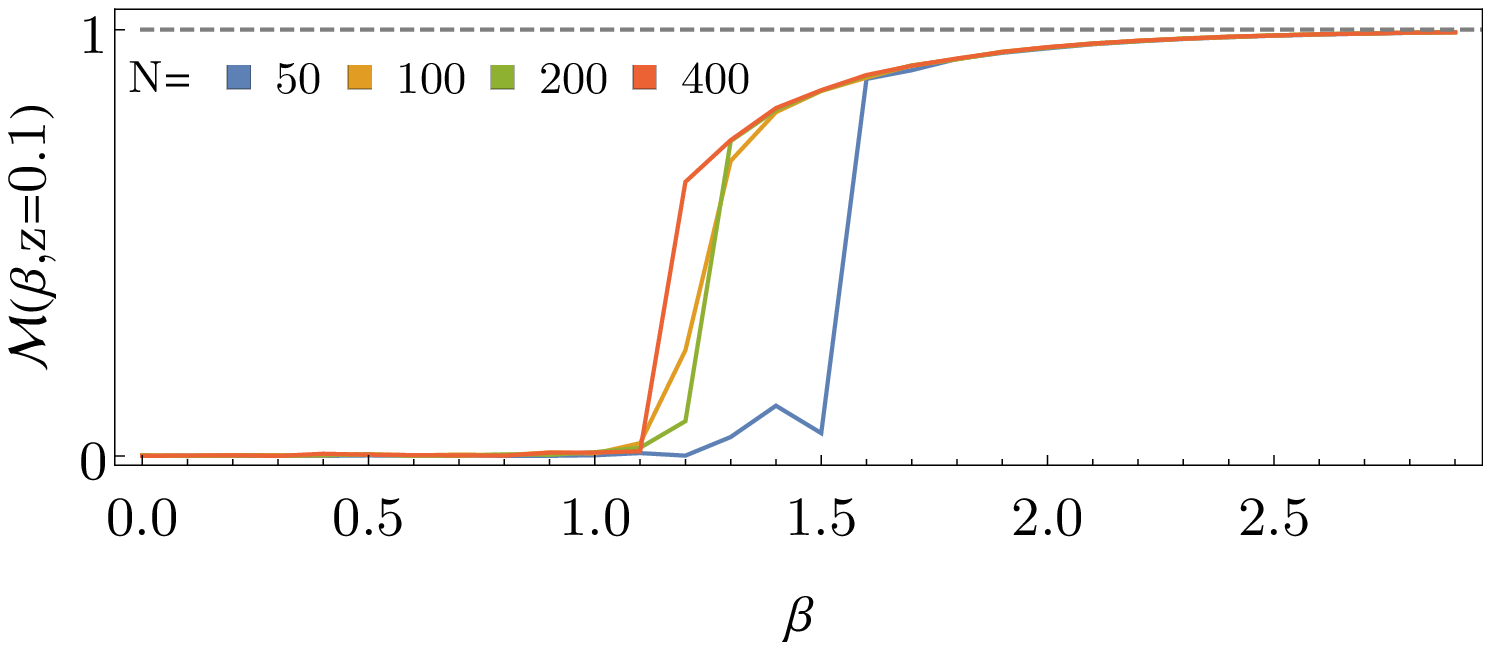}
\includegraphics[width=7.7cm]{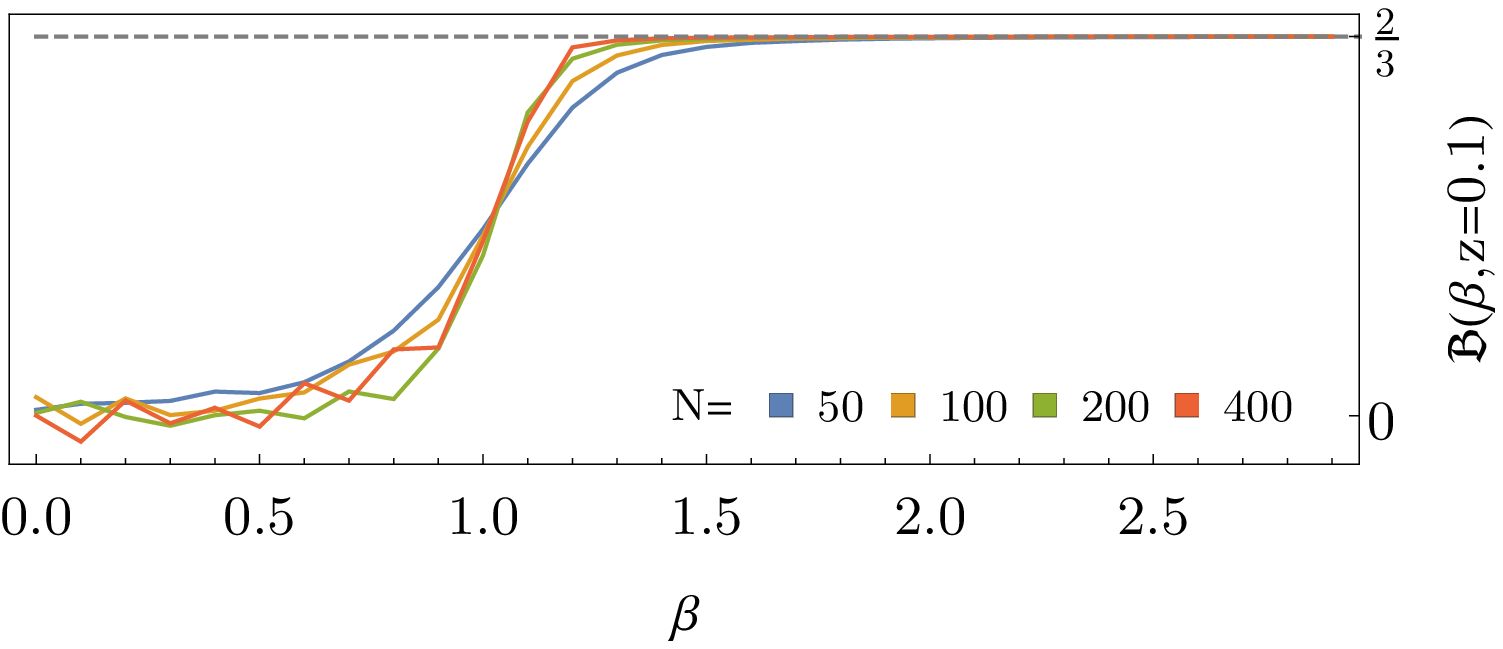}

\includegraphics[width=7.7cm]{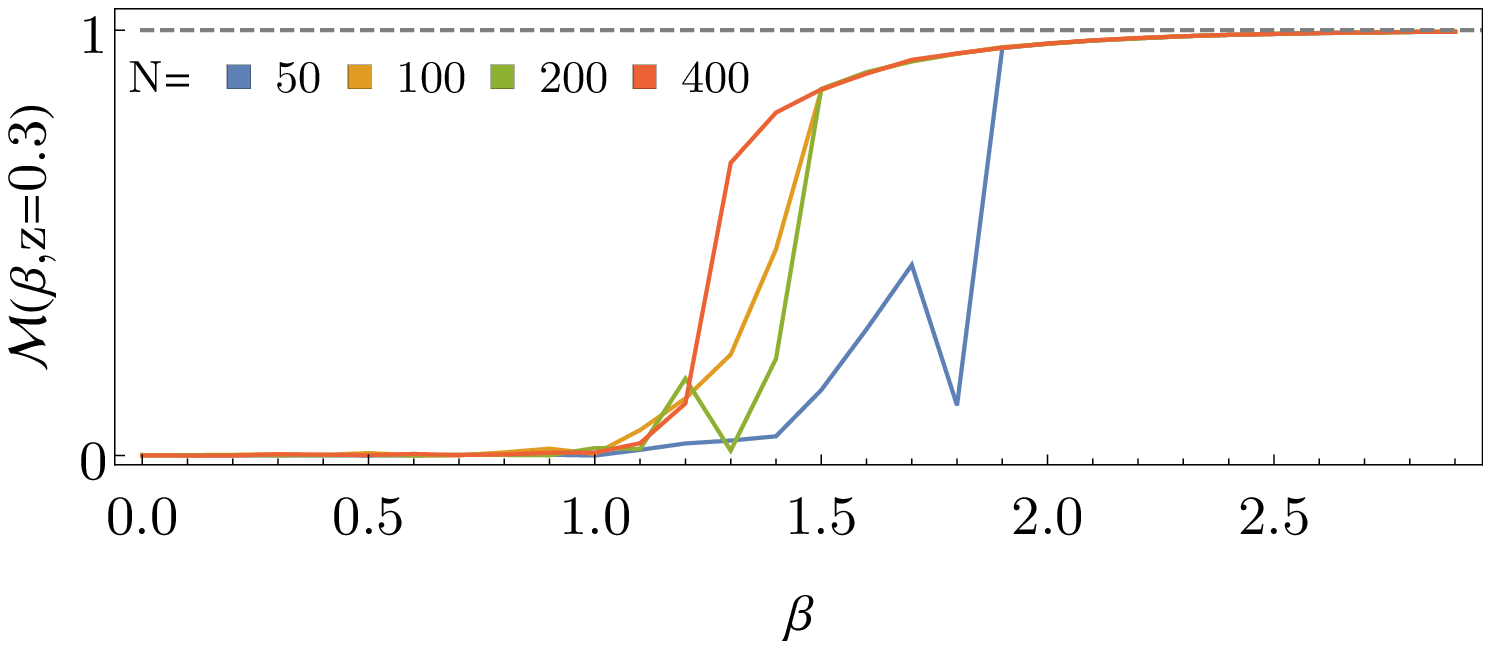}
\includegraphics[width=7.7cm]{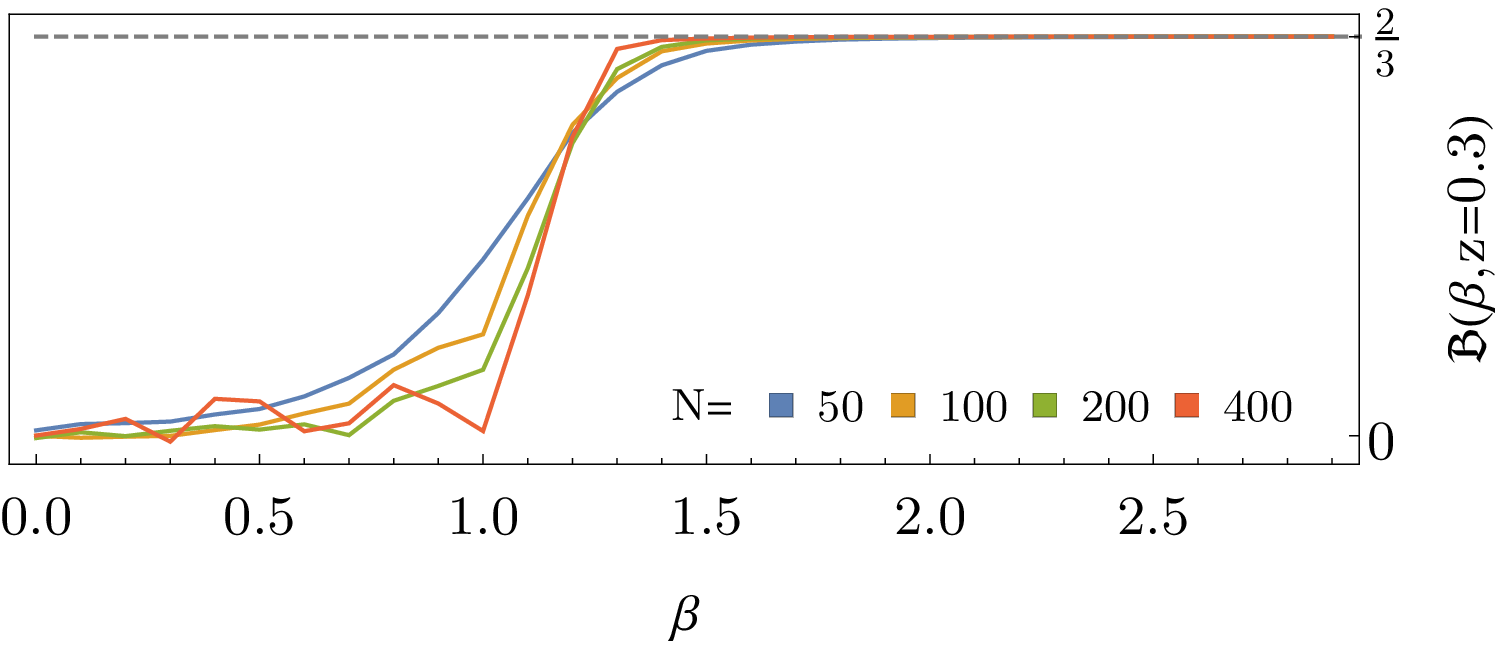}

\includegraphics[width=7.7cm]{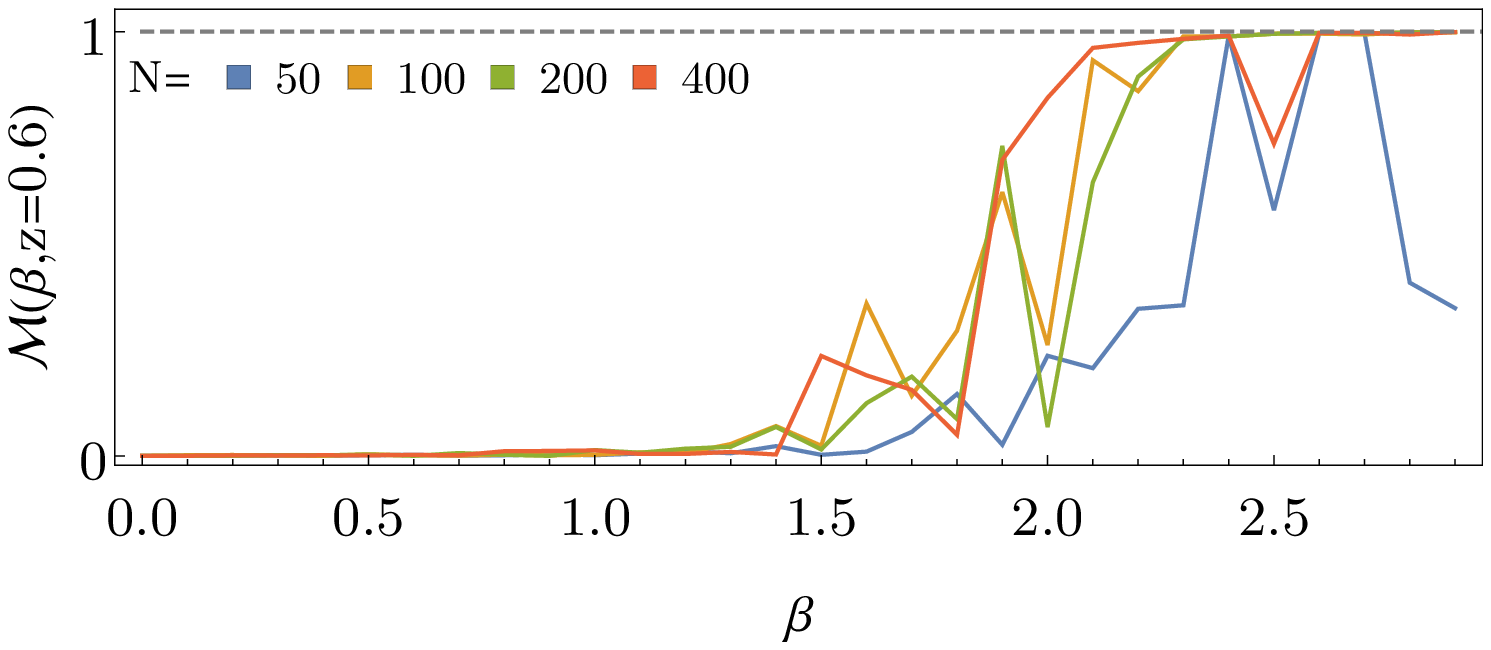}
\includegraphics[width=7.7cm]{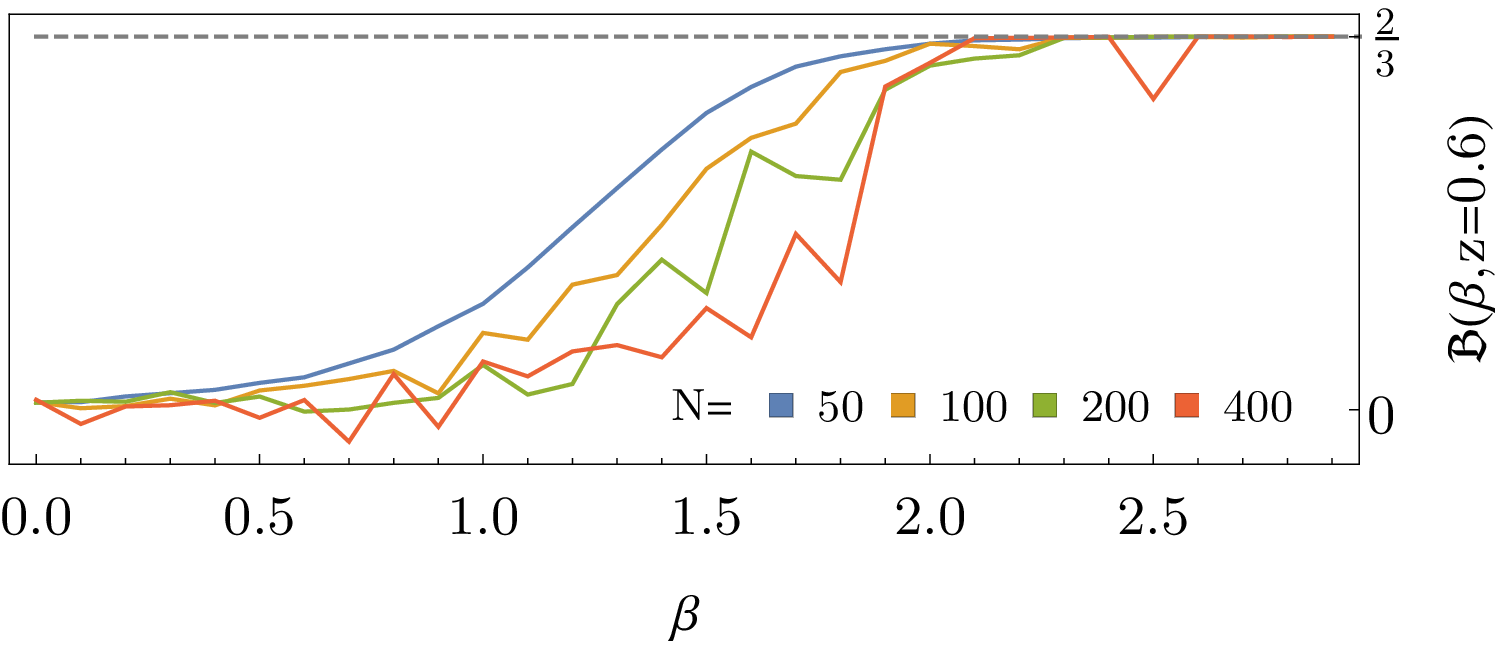}
\caption{Columns - average magnetization $\mathcal{M}$ and Binder parameter $\mathcal{B} $ 
as a function of inverse temperature $\beta$ for different values of $z$ (rows).}
\label{PTPlots}
\end{figure}

Numerical simulations of the system under consideration were performed using regular Monte Carlo 
techniques. 
For any choice of $z=[0,1]$, number of spins $N$ and inverse temperature $\beta :=1/T$ the system was properly thermalized and certain observables were computed. Averaging  $\langle \dots \rangle$ of quantities in question was performed over equilibrium configurations. 
In particular, we have focused our analysis on computing the magnetization
\begin{equation}
s = \frac{1}{N} \sum_{i=1}^N s_i,
\end{equation}
based on which, values of averaged magnetization
\begin{equation}
\mathcal{M}= \langle s \rangle,
\end{equation}
and Binder cumulant 
\begin{equation}
\mathcal{B}=1- \frac{\langle s^4 \rangle }{3\langle s^2 \rangle^2 }.
\end{equation}
were determined. Exemplary plots are shown in Fig. \ref{PTPlots}. Properties of the Binder parameter imply $\mathcal{B} \rightarrow 2/3 \approx 0.667 $ 
in the symmetry broken phase and $\mathcal{B} \rightarrow 0$ in the symmetric phase. Furthermore, at 
the critical point  $0 < \mathcal{B} < 2/3 $.

\begin{figure}[ht!]
\centering
\includegraphics[width=10cm]{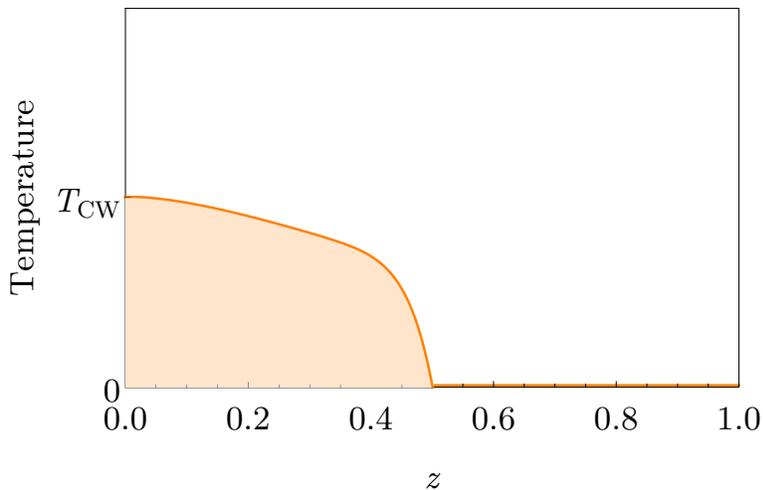}
\caption{Phase diagram of the considered spin system at the Temperature-$z$ plane. 
The $T_{CW}$ corresponds to critical temperature of Curie-Weiss model. The critical 
point extends to a critical line which ends at $z=0.5$. The shadowed region corresponds 
to the symmetry broken phase. In the remaining part of the phase diagram the system 
is in symmetric phase.} 
\label{PTPlot}
\end{figure}

In accordance with the analytical consideration, in the ultralocal limit ($z\rightarrow 0$) the Curie-Weiss 
critical point has been identified.  In the normalized units, location of the critical point is consistent 
with the theoretically predicted value $\beta_{\text CW} =1$ or $T_{\text CW} =1$. Together with the increase
of $z$ the location of the critical point shifts to the smaller temperatures (higher values of $\beta$).
However, while the value of $z=0.5$ is approached, fluctuations are becoming dominant and analysis 
of the Binder cumulant does not support existence of the symmetry broken phase for 
$z>0.5$. This is is accordance with the results of analytical investigations performed in the previous 
subsection. Furthermore, for small values of spin chain size ($N$) finite-size effects discussed in 
the previous subsection skew the results, whereas for large values $N$, the equilibrium state was difficult 
to obtain near the disappearing ($z>1/2$) phase transitions. Based on the joint numerical and analytical 
studies the expected phase space of the system has been sketched in Fig. \ref{PTPlot}.  

One can conclude that for $z\in [0,0.5)$ there exist symmetry broken phase at temperatures 
$T>0$. The associated critical line matches with the Curie-Weiss critical point at $z=0$. 
Furthermore at $z\in [0.5,1]$ only the symmetric phase exists. The result is in accordance 
with the known behavior of the nearest neighbor Ising model on $S^1$, which is realized 
at $z=1$. Worth stressing is that the presented model provides interesting interpolation between 
the two analytical spin systems, one with and second without critical behavior. Therefore, 
the presented model may possibly have relevance outside of the considered here gravitational 
context. More robust analysis, including its quantum counterpart and $h\neq 0$ case is beyond 
the scope of this article and will be pursued in the forthcoming papers. 

\section{Cosmological implications} \label{Cosmo}

In this section we are going to discussion potential cosmological significance of the 
process of geometrogenesis discussed in this article, which would be associated with 
transition in time from $z=0$ to $z=1$.    

Let us begin with emphasizing that one of the most important properties of the 
state of the early universe is its nearly scale-invariant nature of cosmological 
inhomogeneities, reflected in the anisotropy of the Cosmic Microwave Background 
(CMB) radiation. A perfect scale-invariance would mean that the connected two-point 
correlation is invariant 
under rescaling ${\bf x} \rightarrow b {\bf x}$:
\begin{equation}
\langle \phi({\bf x})\phi({\bf y}) \rangle_c = \langle \phi(b{\bf x})\phi(b{\bf y}) \rangle_c,  
\end{equation} 
which means that the \emph{correlator} is a constant function of the distance $r:= |{\bf x}-{\bf y}|$.
The two-point function can be expressed in terms of the so-called power-spectrum 
$\mathcal{P}(k)$ in the following way:
\begin{equation}
\langle \phi({\bf x})\phi({\bf y}) \rangle_c  = \int_{0}^{\infty} \frac{dk}{k} \mathcal{P}(k) \frac{\sin (kr)}{kr}.
\label{twopoint}
\end{equation} 
A simple analysis implies that scale-invariance leads to the constant power spectrum 
$\mathcal{P}(k)=$ const., which is called the Harrison-Zeldovich power spectrum. The 
departure from the scale invariance is parametrized by the spectral index $n$, 
such that  $\mathcal{P}(k) \propto k^{n-1}$. The most up to date observations
of the scalar perturbations indicated that $n_S = 0.9616 \pm 0.0094$ \cite{Ade:2013zuv}. 

In the crumpled phase ($z\rightarrow 0$) represented by the connected graph 
the two point correlation function  $\langle s_i s_j \rangle_c = \langle s_i s_j \rangle- 
\langle s_i \rangle \langle s_j \rangle$ does not depend on the values of $i$ and $j$ 
since all spins are treated at equal footing. The two point correlation function, 
therefore, does not depend on the distance between given points $i$ and $j$ 
(all distances are equal) but only on parameters of the model and temperature. 
However, due the effective change of dimensionality, we cannot directly conclude 
if the fluctuations in the ultralocal state ($z\rightarrow 0$) are scale invariant.
However, some investigations suggest that critical behavior associated  with 
geometrogenesis may lead to nearly scale-invariant powers spectrum of 
primordial perrurbations \cite{Magueijo:2006fu}.  

More detailed analysis of the vacuum quantum fluctuations employees the
following mode equation \cite{Mukohyama:2010xz}
\begin{equation}
\frac{d}{d\eta^2} f_k+\left( k^{2z}-\frac{z_i''}{z_i}\right)f_k=0,    
\end{equation}
where $\eta$ is the conformal time and the $\Delta_z$ Laplace operator (\ref{Deltaz}) has 
been taken into account.  The form of $z_i$ depends on whether tensor $i=T$ or scalar $i=S$
perturbations are considered. With the use of Bunch-Davies type of vacuum normalization, 
the vacuum power spectrum can be expressed as
\begin{equation}
\mathcal{P}(k) := \frac{k^3}{2\pi^2} \left| \frac{f_k}{z_i} \right|^2 = \frac{k^3}{2\pi^2}\frac{1}{2\omega}\frac{1}{z_i^2},
\end{equation}
where dependence of the $\omega$ on $k$ is given be an appropriate dispersion relation. 
In the case considered in this article, the modified spatial part of the Laplace operator 
(\ref{Deltaz}) implies that $\omega^2 = k^{2z}$. This leads to the expression $\mathcal{P}(k) 
\propto k^{3-z}$, such that the scale-invariance is recovered at the Lifshitz point 
($z=3$). In turn, the ultralocal limit leads to the cubic power spectrum $\mathcal{P}(k) 
\propto k^{3}$ describing white noise type of fluctuations for which the two-point 
correlation function is just vanishing:
\begin{equation}
\langle \phi({\bf x})\phi({\bf y}) \rangle_c = 0. 
\end{equation} 
This is in agreement with the the analysis of the ultralocal state performed in the 
context of loop quantum cosmology \cite{Mielczarek:2014kea}. Moreover, 
worth mentioning is that the white noise power spectrum is also scale-invariant, 
though, in a trivial way since the correlation function is just equal zero.  

The possibility interesting from the cosmological perspective is the transition across 
the second-order transition occurring in the setup discussed in the previous section. 
Since such transition is hypothesized to occur in a finite amount of time, formation 
of domains as well as associated topological defects are expected. The process is 
described by the so-called Kibble-Zurek \cite{Kibble:1976sj,Zurek:1985qw} mechanism which takes 
into account scaling of both correlation length: 
\begin{equation}
\xi = \xi_0 \left|1-T/T_C\right|^{-\nu}, 
\end{equation}
and the relaxation time 
\begin{equation}
\tau= \tau_0 \left|1-T/T_C\right|^{-\nu Z},
\end{equation}
where $\nu$ and $Z$ are critical exponents characterizing the \emph{equilibrium} version 
of the considered phase transition. Based on this, the diameter of the relaxation horizon 
can be expressed as follows: 
\begin{equation}
\xi_c = \int_0^{t}  \frac{\xi(T)}{\tau(T)}dt \propto \xi_0\frac{\tau_Q}{\tau_0} \left(1-T/T_C\right)^{\nu(Z-1)+1},   
\end{equation}
where we introduced the so-called quench time $\tau_Q$  which characterizes 
a speed of the transition.  The process of relaxation will stop when the size of the  
relaxation horizon becomes equal with the correlation length $\xi_c(T_F)=\xi(T_F)$.
This condition leads to the following expression for the domain size:  
\begin{equation}
\xi \approx \xi_0 \left(\frac{\tau_Q}{\tau_0}\right)^{\frac{\nu}{\nu Z+1}}. 
\end{equation}
The topological defects are formed at the boundaries between the domains. 
For  typical values of the parameters we obtain $ \xi\approx l_{\rm Pl}$. 
Assuming that there is no inflation the rough estimation of the density of defects is:
\begin{equation}
n \sim  \frac{1}{\xi^3} \left(\frac{T_{\rm CMB}}{T_{\rm Pl}}\right)^3 \sim 1 \frac{\text{defect}}{\text{mm}^3}.
\end{equation}   
If the phase of inflation occurs after the second order phase transition, the value 
of $n$ can be further diluted with the $e^{-3N}$ factor, where $N$ is the number 
of inflationary $e$-folds. This number is expected to not be less that $N\sim 60$.
In consequence, the expected present number of defects is suppresses by at least 
the enormous $10^{-180}$ factor, which makes any current empirical relevance of 
the defects groundless.  

\section{Summary and discussion} \label{Summary}

The main purpose of this article was to investigate interconnection between the 
phase of ultralocality and the process of geometrogenesis. The studies were 
inspired by the observation made in Ref. \cite{Mielczarek:2017cdp} which suggested 
that the ultralocality shares some features of the crumpled state of gravitational 
field characterized by enormous level of connectivity. 

In our studies, the anisotropic scaling introduced in the context of Horava-Lifshitz 
theory has been considered. In such a case, the ultralocal limit is recovered by 
taking the $z\rightarrow 0$ limit, while the standard relativistic scaling is satisfied 
at $z=1$.  By analyzing the properties of the Laplace operator, connectivity structure 
in the $z\rightarrow 0$ has been reconstructed. We proved analytically than in the 
$z\rightarrow 0$ limit the spatial connectivity is represented by a complete graph.
However, such relation is satisfied for finite system only. We have shown that 
when the number of nodes $N$ grows to infinity, the complete graph disintegrates into 
a set of points. This observation supports the discussion of Ref. \cite{Oriti:2018tym}, 
which stressed the dependence of the semiclassical limit on the 
number of degrees of freedom. In our case, we observed that depending on whether 
finite or infinite number of nodes is considered, the corresponding phase may have 
qualitatively different properties.  

Furthermore, we have shown that the transition between the ultralocal state and 
the geometric state of gravity can be associated with the process of removing 
links from the graph representing connectivity structure between the points of 
space. The process can describe hypothetical geometrogenesis occurring 
in the early universe. In such a picture, the evolving (maturing) Universe drastically 
reduces number of spatial interconnections. Worth mentioning is that there are 
also other complex systems observed in Nature, which exhibit such reduction 
of connectivity. One of the seminal examples of such a behavior is the so-called 
\emph{synaptic pruning} occurring in the maturing brain \cite{Pruning}. 
While newborn brain is full of connections, its maturing reduces relative number 
of connections with respect to the number of neurons (nodes of a graph). Such 
reduction is associated with both optimization of the brain functioning and the 
process of learning. From such a perspective one can hypothesize that the maturing 
universe removes some irrelevant connection in its structure. In consequence,
the connectivity structure (encoded in the weights $w_{ij}$) undergoes specialization 
in similarity with the process of learning. 

In order to study a potential critical behavior associated with the geometrogenesis
we have introduced Ising type matter defined at the nodes of the considered graph. 
Both analytical and numerical studies of the system were performed for $z\in[0,1]$, with the
analysis limited to the special case of a ring graph ($S^1$ topology) 
at $z=1$.  As we have shown, the Curie-Weiss critical point is recovered at $z=0$
and extends as a critical line in the  $z\in[0,0.5)$ range. The second order nature 
of the transition line has been confirmed by the numerical investigations of the Binder 
cumulant. We have shown that the symmetry broken phase does not occur in the 
$z\in(0.5,1]$ range. 

Finally, some possible cosmological consequences of the considered scenario were
investigated. In particular, we have confirmed, that the ultralocal limit is associated 
with the white noise type of fluctuations. Furthermore, we have stressed that the 
identified critical line may lead to the formation of domains and topological defects. 
Observational constraints on the number of topological defects may allow one
to put some constraints on the discussed features in early universe.  
  
\ack
We would like to thank dr Jakub Prauzner-Bechcicki for helpful suggestions.
Authors are supported by the Grant DEC-2014/13/D/ST2/01895 of the National 
Centre of Science. JM is supported by the Mobilno\'s\'c Plus Grant 1641/MON/V/2017/0 
of the Polish Ministry of Science and Higher Education. 

\appendix{
\section{Proof of real-valuedness of $\Delta_z$ }
\label{realMatrixAppendix}
This property can be formally proven with the following calculation:
\begin{align}
&\Im \left[\left( \Delta_z \right)_{mn}\right] =  \frac{1}{N}\sum_{k=1}^{N}  
\sin \left[2\pi (k-1)(m-n)/N\right] 2^z\left[1-\cos\left( 2\pi \frac{(k-1)}{N} \right) \right]^z \nonumber \\ 
&= \sum_{r=0}^{\infty} (-1)^r \left( \begin{array}{c} z \\ r \end{array} \right) \frac{2^z}{N}  
\sum_{k=1}^{N}  \sin \left[2\pi (k-1)(m-n)/N\right] \cos^r \left[2\pi (k-1)/N\right],
\end{align}
which followed by the Fourier decomposition, 
\begin{equation}
\cos^r \left[2\pi (k-1)/N\right] =  \sum_{l=1}^{N} c_l \cos \left[2\pi (k-1) l /N\right]
\end{equation}
together with the orthogonality relation
\begin{equation}
\sum_{k=1}^{N} \sin \left[2\pi (k-1)m/N\right]\cos \left[2\pi (k-1)n/N\right] = 0,      
\end{equation}
completes the proof that $\Im \left[\left( \Delta_z \right)_{mn}\right] =0$.

\section{Intuition behind Ruelle's condition and it's modification}
\label{ModifiedRuelleAppendix}

The physical motivation behind Ruelle's condition is that energy of interaction between two infinite domains 
of $\dots +++++++ \dots$ and $\dots ------- \dots$ orientations of spins must be infinite to withstand thermal 
fluctuations that could spoil stability of a given phase. In the case of the Ising model on $\mathbb{R}$ there 
is only a single boundary between the domains. And there is a single contribution to the interaction from $J(1)$, 
there are two pairs which interact with $J(2)$, three pairs which interact with $J(3)$, etc. leading to the expression 
(\ref{M1}). In the considered case of $S^1$ topology of the graph for $z=1$, instead of a single boundary there 
are two boundaries between the domains, which have to be taken into account while calculating the energy 
of interactions between the domains. Before, we study the general case let us calculate first the energy of 
interaction between the domains for $z=0$ and $z=1$. For $z=0$ there is a Curie-Weiss phase transition, so 
we expect that $M_1=\infty$, on the other hand there is no phase transition for $z=1$ in the case of considered 
ring graph, and in consequence $M_1 < \infty$. Let us suppose that we have two domains each containing 
$N/2$ Ising spins. Then there are $\sim \left(\frac{N}{2}\right)^2$ links between the two domains each associated 
with the same coupling $J(n)=\frac{1}{N}$. In consequence, the energy of interaction between the two domains is 
$M_1 = \left(\frac{N}{2}\right)^2 \frac{1}{N} \sim N$, which is infinite in the thermodynamical limit ($N\rightarrow \infty$). 
On the other hand at $z=1$ only $J(1)$ contributes and energy of interaction between the two domains
is just $2 J(1) = 2$, which is finite even in the thermodynamic limit.

Assuming that $N$ is even, the interaction energy between the two domains (each of the $N/2$ size) 
defined at the ring graph can be calculated as follows. Let us suppose that $i=1,2, \dots,  N/2$ are "+" 
spins and  $i=N/2, N/2+1, \dots,  N$ are "-", the interaction energy between the $i$-th positive spin 
and all spins from the domain of negative spins is: 
\begin{equation}
E_i = \sum_{n=N/2}^{N-1}J(n-(i-1)).
\end{equation} 
Based on this, $M_1$ can calculated by summing up $E_i$ over all of the spins belonging to the domains 
of the positive spins, i.e. 
\begin{align}
\label{M1_expl}
M_1 = \sum_{i=1}^{N/2} \sum_{n=N/2}^{N-1}J(n-(i-1)) = \sum_{k=1}^{N} a_k,
\end{align}   
where 
\begin{equation}
a_1 = - 2^{2(z-1)} N \lim_{k\rightarrow 1}  \left[ \sin \left( \pi \frac{k-1}{N} \right)\right]^{2 z},
\end{equation}
and for $k>1$ 
\begin{equation}
a_k = \frac{2^{2(z-1)}}{N} \left( 1-(-1)^{k-1} \right))^2 \left[ \sin \left( \pi \frac{k-1}{N} \right)\right]^{2(z-1)}. 
\end{equation}
In Fig. \ref{M1_N20} we plot $M_1$ as a function of $z$ for the $N=20$. As expected, the function converges 
to $M_1=2$ for $z=1$ and is equal to $\left(\frac{N}{2}\right)^2 \frac{1}{N} = \frac{N}{4}$ in the ultralocal limit. 
\begin{figure}[ht!]
\centering
\includegraphics[width=10cm]{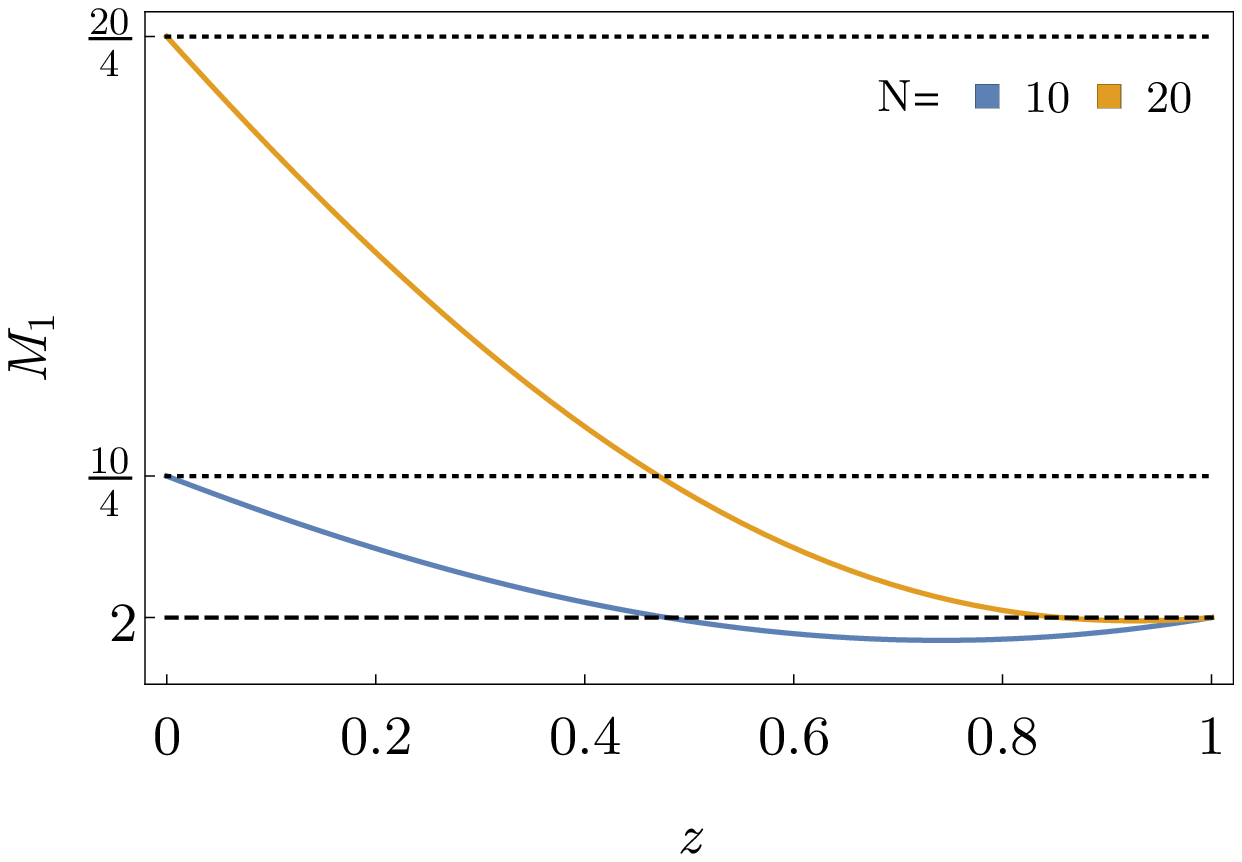}
\caption{Values of function $M_1(z)$ for finite-size ($N$) spin chains. The curves correctly match with 
$M_1=N/4$ at $z=0$ and  $M_1=2$ at $z=1$.}
\label{M1_N20}
\end{figure}   
In order to verify what is the behavior of $M_1$ for $N\rightarrow \infty$, we can study 
convergence property for the sum $M_1(N,z)$ which can be further simplified to 
\begin{align}
M_1 = a_1+\sum_{m=1}^{N/2} b_m,
\end{align}   
where 
\begin{equation}
b_m = \frac{2^{2z}}{N} \left[ \sin \left( \pi \frac{2m-1}{N} \right)\right]^{2(z-1)}. 
\end{equation}

\subsection{Alternative method of calculating Ruelle's condition}
\label{AlternativeRuelleAppendix}
Simultaneously, another interesting insight can be given. One could argue that in the 
thermodynamic limit the expression (\ref{Jnthermo}) has lost the information about the 
ring topology ($J(N/2 -1) \neq J(N/2 +1) $) and that one could ``recover" the right 
behavior by calculating $M_1$ as
\begin{equation}
\label{M1solution2}
\lim_{N \rightarrow \infty} \sum_{n=1}^{N} n J(n/2),
\end{equation}
which in fact gives the same result\footnote{Actually, the exactness of this expression is 
lost at single point of $z=0$.} as (\ref{M1solution}), even though the input for $J$ was 
assumed to be an integer.

\subsection{Finite size effects}
\label{FiniteSizeAppendix}
Skimming through the presented analysis one might get the idea that numerical simulations of the 
presented spin system might be a viable option. This presumption however, is true only for large 
spin systems (assuming we can simulate such) and $z<1/2$, in rest of the cases we can predict 
finite-size effects simply due to ring topology. One can see this most clearly by working through the sum (\ref{M1_expl}):
\begin{equation}
\begin{split}
\label{JnSum}
& \sum _{i=1}^{N/2} \sum _{n= N/2}^{N-1} J (n-(i-1)) =\sum _{i=1}^{N/ 2} \sum _{n=1}^{N/2} J \left(-(i-1)+n+ N/2-1\right)
\\ & \overset{n-i=k}{=} \sum_{k=1}^{N/2} 2 \left(\frac{N}{2}-k\right) J \left(\frac{N}{2}-k\right)
+\frac{N}{2} J \left(\frac{N}{2}\right) =\sum _{u=1}^{N/2-1} 2 u J(u)+\frac{N}{2} J \left(\frac{N}{2}\right)\end{split}.
\end{equation}
The second term of the last expression holds the information about the interaction of each spin 
with its partner from the opposite side of the ring. It can be shown that this term vanishes in the 
thermodynamic limit, however it is worthwhile to look at its dependence on $N$ (Fig. \ref{N2JN2}), 
to notice the mentioned finite-size effects, strongest near $ z \approx 1 $. Nevertheless, numerical 
results can provide us some qualitative information about the transitions of the critical temperatures.
\begin{figure}[ht!]
\centering
\includegraphics[width=10cm, angle = 0]{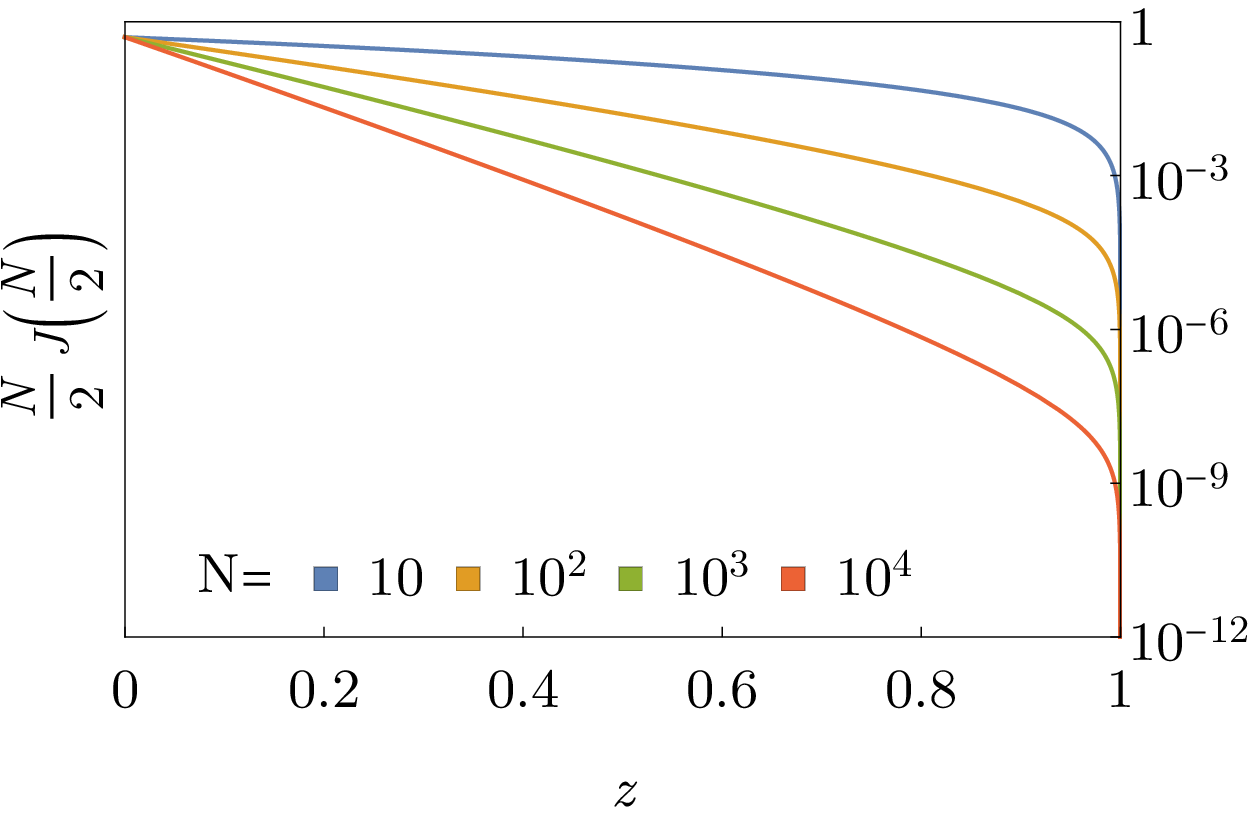}
\caption{Contribution of the far-away interaction to $M_1(z)$.}
\label{N2JN2}
\end{figure}   

}

\end{document}